\documentclass[conference]{IEEEtran}

\usepackage{epsfig,endnotes}
\usepackage{dsfont}
\usepackage{times}
\usepackage{balance}
\usepackage{color}
\usepackage{ae, aecompl}
\usepackage{graphicx}
\usepackage{amsfonts}
\usepackage{amsmath}
\usepackage{mathrsfs}
\usepackage{mathtools}
\usepackage{subfigure}
\usepackage{wrapfig}
\usepackage{xypic}
\usepackage{multirow}
\usepackage{enumerate}
\usepackage[font={small}]{caption}
\usepackage{balance}
\usepackage{amsmath}
\usepackage{verbatim}
\usepackage{xspace}
\usepackage{listings}
\usepackage{textcomp}
\usepackage{algpseudocode,algorithm,algorithmicx}
\algrenewcommand\algorithmicrequire{\textbf{Precondition:}}
\algrenewcommand\algorithmicensure{\textbf{Postcondition:}}
\algnewcommand\algorithmicswitch{\textbf{switch}}
\algnewcommand\algorithmiccase{\textbf{case}}
\algnewcommand\algorithmicassert{\texttt{assert}}
\algnewcommand\Assert[1]{\State \algorithmicassert(#1)}%
\algdef{SE}[SWITCH]{Switch}{EndSwitch}[1]{\algorithmicswitch\ #1\ \algorithmicdo}{\algorithmicend\ \algorithmicswitch}%
\algdef{SE}[CASE]{Case}{EndCase}[1]{\algorithmiccase\ #1}{\algorithmicend\ \algorithmiccase}%
\algtext*{EndSwitch}%
\algtext*{EndCase}%
\algnewcommand{\algorithmicgoto}{\textbf{go to}}%
\algnewcommand{\GoTo}[1]{\algorithmicgoto~\ref{#1}}%

\newcommand{\storageget}[1]{\textsc{Get}(#1)}
\newcommand{\storageput}[1]{\textsc{Put}(#1)}

\usepackage{paralist}
\renewcommand{\itemize}{\compactitem}
 \renewcommand{\enumerate}{\compactenum}

\newcommand{\our}{\text{ReplicaTEE}\xspace}
\newcommand{\layer}{\text{EML}\xspace}
\newcommand{\cloud}{${C}$\xspace}
\newcommand{\storage}{\text{BSL}\xspace}

\newcommand{\verylongrightarrowRF}[1]{\xymatrix@1{{\hole} \ar[rr]^{{#1}} && {\hole}}}
\newcommand{\verylongrightarrowRFQ}[1]{\xymatrix@1{{\hole} \ar@{.>}[rr]^{{#1}} && {\hole}}}
\newcommand{\verylongleftarrowRF}[1]{\xymatrix@1{{\hole} && {\hole} \ar[ll]_{{#1}} }}
\newcommand{\verylongleftarrowRFQ}[1]{\xymatrix@1{{\hole} && {\hole} \ar@{.>}[ll]_{{#1}} }}
\newcommand{\verylongrightarrowUS}[1]{\xymatrix@1{{\hole} \ar@{~>}[rr]^{{#1}} && {\hole}}}
\newcommand{\verylongleftarrowUS}[1]{\xymatrix@1{{\hole} && {\hole} \ar@{~>}[ll]_{{#1}} }}

\newcommand{\verylongrightarrowRFM}[1]{\xymatrix@1{{\hole} \ar[rr]|(.45){{#1}} && {\hole}}}
\newcommand{\verylongleftarrowRFM}[1]{\xymatrix@1{{\hole} && {\hole} \ar[ll]|{{#1}} }}

\newcommand{\verylongrightarrowST}[1]{\xymatrix@1{{\hole} \ar@2{~>}[rr]^{{#1}} && {\hole}}}
\newcommand{\verylongleftarrowST}[1]{\xymatrix@1{{\hole} && {\hole} \ar@2{~>}[ll]_{{#1}} }}

\begin{document}

\title{ReplicaTEE: Enabling Seamless Replication of SGX Enclaves in the Cloud}
%\author{Claudio, Ghassan, Wenting, Sergey}
\author{
\IEEEauthorblockN{Claudio Soriente}
\IEEEauthorblockA{NEC Laboratories Europe \\
claudio.soriente@emea.nec.com}
\and
\IEEEauthorblockN{Ghassan Karame}
\IEEEauthorblockA{NEC Laboratories Europe \\
ghassan.karame@neclab.eu}
\and
\IEEEauthorblockN{Wenting Li}
\IEEEauthorblockA{NEC Laboratories Europe \\
wenting.li@neclab.eu}
\and
\IEEEauthorblockN{Sergey Fedorov}
\IEEEauthorblockA{NEC Laboratories Europe \\
sergey.fedorov@neclab.eu}
}

\maketitle

\begin{abstract}
With the proliferation of Trusted Execution Environments (TEEs) such as Intel SGX, a number of cloud providers will soon introduce TEE capabilities within their offering (e.g., Microsoft Azure). Although the integration of SGX within the cloud considerably strengthens the threat model for cloud applications, the current model to deploy and provision enclaves prevents the cloud operator from adding or removing enclaves dynamically---thus preventing elasticity for TEE-based applications in the cloud.

In this paper, we propose \our, a solution that enables seamless provisioning and decommissioning of TEE-based applications in the cloud. \our leverages an SGX-based provisioning layer that interfaces with a Byzantine Fault-Tolerant storage service to securely orchestrate enclave replication in the cloud, without the active intervention of the application owner.
Namely, in \our, the application owner entrusts application secret to the provisioning layer; the latter handles all enclave commissioning and de-commissioning operations throughout the application lifetime. We analyze the security of \our and show that it is secure against attacks by a powerful adversary that can compromise a large fraction of the cloud infrastructure. We implement a prototype of \our in a realistic cloud environment and evaluate its performance. \our moderately increments the TCB by $\approx 800$ LoC. Our evaluation shows that \our does not add significant overhead to existing SGX-based applications.
\end{abstract}

\section{Introduction}
\label{sec:intro}
In the last few years, the cloud has been gaining several adopters among SMEs and
large businesses that are mainly interested in minimizing the costs of deployment, management, and maintenance of their computing infrastructure. Cost effectiveness is realized in the cloud by coupling multi-tenancy with tailored distributed algorithms that ensure unprecedented levels of scalability and elasticity at low costs~\cite{ArmknechtBKY15}.

With the recent proliferation of Trusted Execution Environments (TEEs) such as Intel SGX, a number of cloud providers will soon introduce TEE capabilities within their offering (e.g., Microsoft Azure~\cite{azure}). The embedding of TEEs within the cloud allows the design of secure applications that can tolerate malware and system vulnerabilities, as application-specific secrets are shielded from any privileged code on the same host. As such, SGX has fueled innovation in the area of secure computation, with an increasing number of proposals that promote TEE-based applications in the cloud~\cite{krawiecka18www,shih16sdnnfv,schuster15sp}.

Although the integration of SGX within the cloud considerably strengthens the threat model for cloud applications, the current model to deploy and provision an enclave, prevents the cloud operator from adding or removing enclaves dynamically---thus effectively hampering elasticity for TEE-based applications in the cloud. Namely, SGX enclaves bear no secrets when deployed; secrets are securely provisioned to the enclave by the application owner (also known as Independent Software Vendor or ISV) after he attests the application code and makes sure that it runs untampered in an enclave on an SGX-enabled platform. In a nutshell, dynamic enclave allocation for TEE-based applications in the cloud requires the ISV to be online throughout the whole application lifetime. The only alternative for an ISV is to entrust the secrets of his application to the cloud provider (in a way similar to the provisioning of Virtual Machine images that carry secret material). This, however, obviates the shift to deploy SGX enclaves in the cloud since it exposes all application secrets to malware that may potentially penetrate the cloud infrastructure.%the cloud provider.

Although the community features a number of studies on SGX security in the cloud~\cite{matetic17sec,brandenburger17dsn,206170}, no previous work has addressed the problem of enabling seamless provisioning and decommissioning of enclaves in the cloud. Here, there are a number of challenges to overcome. One the one hand, such a service should remove the need of an online ISV. On the other hand, it should warrant ISVs the same security provisions of the current deployment and provisioning models, where ISVs attest and provision secret material to their applications. Furthermore, unrestricted enclave replication in the cloud may amplify the effectiveness of \emph{forking attacks} for application that keep persistent state~\cite{brandenburger17dsn}. In a forking attack, the adversary runs several instances of an application and provides them with different state or inputs in order to influence their behavior. For example, consider an authentication service running in SGX enclaves. To mitigate brute-force attacks, the service may use rate-limiting and, for example, allow up to $3$ password trials per account. An adversary that manages to compromise the cloud infrastructure could launch several instances of the service in order to increase the number of trials per account and brute-force passwords. A service that automatically provisions enclaves must, therefore, control the number of running enclaves for a given application at all times, despite potential malware that may penetrate the cloud infrastructure.

In this paper, we propose \our, a solution that enables dynamic enclave replication and de-commissioning for TEE-based applications in the cloud. \our leverages a distributed SGX-based service layer that interfaces with a Byzantine Fault-Tolerant (BFT) storage layer to orchestrate secure and dynamic enclave replication in the cloud. Namely, in \our, the ISV entrusts application secrets to the service layer and can go offline. The service layer is a thin software layer that runs in SGX and handles commissioning and de-commissioning of enclave replicas on behalf of the ISV. Application secrets are, therefore, shielded away from malware that penetrates the cloud, as they are securely transferred from the ISV to the service layer onto application enclaves. The service layer also controls the number of running replicas for a given application, in order to mitigate forking attacks against victim applications. Finally, in order to prevent forking attacks to the service layer itself, \our uses a distributed BFT storage layer that guarantees dependable storage despite compromise of a fraction of its nodes.

We design \our to be fully compliant with the existing Intel SGX SDK. We analyze the security of \our and show that it enables secure enclave provisioning and decommissioning even in presence of a powerful adversary that compromises a large fraction of the cloud infrastructure. We also implement a prototype of \our in a realistic cloud environment and evaluate its performance. Our evaluation shows that \our only moderately increments the TCB by approximately 800 Lines of Code (LoC) and does not add significant overhead to existing SGX-based applications.

The remainder of this paper is structured as follows. In Section~\ref{sec:background}, we review Intel SGX and BFT storage solutions that leverage TEEs. In Section~\ref{sec:model}, we introduce our system and threat models, we discuss our design goals and provide a brief overview of our solution. In Section~\ref{sec:one}, we present \our and analyze its security. In Section~\ref{sec:impl}, we evaluate a prototype implementation based on the integration of \our with a realistic cloud environment. In Section~\ref{sec:related}, we review related work in the area, and we conclude the paper in Section~\ref{sec:conclusion}.

\section{Preliminaries}
\label{sec:background}
In this section, we briefly overview the main operations of Intel SGX and we outline existing Byzantine Fault-Tolerant storage protocols that leverage TEEs.

\subsection{Intel SGX}
Software Guard Extensions (SGX) is the latest realization of Trusted Execution Environment (TEE) by Intel, available on Skylake and later CPUs. It allows application to run in secure containers called \emph{enclaves} with dedicated memory regions that are secured with on-chip memory encryption. Access to the encrypted memory is mediated by the hardware, effectively excluding the OS or any other software from the Trusted Computing Base (TCB).

Privileged code on the planform can create and add data to an enclave with instructions \texttt{ECREATE}, \texttt{EADD}, \texttt{EINIT}. After creation, the enclave code can only be invoked using a thin interface via instructions \texttt{ENTER} and \texttt{ERESUME}; enclave code returns by calling \texttt{EEXIT}, which  ensures that any sensitive information is flushed before control is given back to the OS.

State persistence across reboots is available through \emph{sealing}, i.e., hardware-managed authenticated and confidential persistent storage. Enclaves can use instructions \texttt{EREPORT} and \texttt{EGETKEY} to retrieve an enclave-specific (and platform-specific) key to encrypt data before writing it on persistent storage. Keys are uniquely bound to the identity of an enclave so that no other software including no other enclave can access them.\footnote{Keys may also be bound to a ``sealing authority'' in order to allow secure storage across different versions of the same application.} Note that the sealing functionality that offers SGX does not ensure freshness. That is, a malicious OS may present stale state information to an enclave, what is commonly referred to as a \emph{rollback attack}~\cite{Strackx14acsac}. This is in part mitigated by the use of monotonic counters provided by the platform. However, monotonic counters are apparently slow and the registries where they are stored wear out with usage~\cite{matetic17sec}.

SGX allows a remote party to verify that a piece of code runs in an enclave on an SGX-enabled platform. This mechanisms, called remote attestation, uses a Direct Anonymous Attestation (DAA)~\cite{brickell04ccs} scheme that provides platform anonymity, i.e., the verifier is assured that the enclave runs on an SGX platform without being able to tell it apart from other SGX platforms. Remote attestation in SGX is a two-step process. During the first step, the enclave to be attested proves its identity to a system enclave present on every platform and called \emph{quoting enclave}. The latter has access to the DAA signing key and produces a publicly verifiable \emph{quote} that allows the verifier to remotely attest the enclave. In its current implementation, attestation involves an Intel service (Intel Attestation Service, IAS) that mediates communication between quoting enclaves and remote verifiers. In particular, the IAS only allows registered parties to issue remote attestation requests. Also, the quote produced by a quoting enclave is encrypted under the IAS public key, so that only the IAS can proceed with the verification. The IAS then signs a publicly verifiable statement to confirm that the enclave runs on an SGX platform.
As a by-product of the attestation protocol, the prover and the verifier establish a mutually authenticated Diffie-Hellman key. In particular, the verifier signs its ephemeral key and the enclave must hold the corresponding verification key to verify the signature. Also, the quoting enclave (and IAS) guarantee that the prover ephemeral key belongs to that specific enclave running on an SGX platform.

\subsection{Byzantine Fault-Tolerant Storage using TEEs}

The community features a large number of Byzantine Fault-Tolerant protocols (BFT)~\cite{BChain,Next700,powerstore} based on state replication across different nodes, called ``replicas''. Some replicas may be faulty and their failure mode can be either {\em crash} or {\em Byzantine} (i.e., deviating arbitrarily from the protocol~\cite{Byzantine}). Classical BFT protocols require $3f+1$ nodes and $O(n^2)$ communication rounds among these nodes in order to tolerate up to $f$ Byzantine nodes.

Since agreement in classical BFT is rather expensive, prior work has attempted to improve performance by leveraging trusted hardware. Namely, previous work showed how to use trusted hardware to reduce the number of replicas and/or communication rounds for
BFT protocols~\cite{Behl2017,CheapBFT,MinBFT}. For example, MinBFT~\cite{MinBFT} is an efficient BFT protocol that reduces the communication rounds and the number of replicas used by conventional BFT protocols, by leveraging functionality from TEEs, such as Intel SGX. As a result, the number of required replicas is reduced from $3f+1$ to $2f+1$.
In MinBFT writers send \emph{write} requests (e.g., using a PUT interface) to the replicas, which are all expected to execute the requests in the same order (i.e., maintain a common state). Readers can read content previously written onto the replica nodes. The main idea of MinBFT is to rely on the sequential monotonic counter provided by trusted hardware, in order to bind each message sent to a unique counter value. This is ensured by requiring a signature from the local TEE on all messages sent by the replica; the intuition is that the TEE will sign messages with a given counter value only once, thereby preventing replicas from assigning the same counter value to different messages---commonly referred to as \emph{equivocation}.  More details about MinBFT can be found in Appendix~\ref{ap:minbft}.

\section{Model \& Overview}\label{sec:model}
In this section, we introduce our solution, \our, which enables seamless replication of TEE-based applications in the cloud. We start by describing our system and threat model.

\subsection{System Model}

We consider a scenario where a cloud provider manages a set of Intel SGX-enabled platforms.
Application owners, also known as Independent Software Vendors (ISV), can upload code to be executed on such platforms. Applications could either run computation on behalf of the ISV such as a map-reduce service~\cite{schuster15sp}, or provide public functionalities such as an online password-strengthening service~\cite{krawiecka18www}.

\noindent\textbf{Deployment.} In a real-world deployment of \our, application owners would acquire (e.g., rent) VMs at the cloud and split the logic of their applications (e.g., by using available tools~\cite{lind17atc}) in sensitive code to be run in an enclave and non-sensitive code that can run inside the VM. Therefore, each of the cloud platforms would host VMs from different tenants and each VM would have one or more enclaves. However, for the sake of simplicity, we assume in this paper that the entire application code is executed in enclaves. Given this assumption, each of the cloud platforms hosts multiple enclaves belonging to different ISVs.

\noindent\textbf{Dynamic Provisioning.} Conforming with current elastic cloud settings, we assume that multiple \emph{instances} of the same application enclave may dynamically be started or shut down. In the following, we use the term \emph{application enclave} to refer to an instance of application code running in an enclave, and we use \emph{application} to denote the logical entity spanning multiple enclaves running the same code.

We are agnostic on how the decision to add or remove application enclaves for a given application is made. For example, this decision may be taken by the cloud for reasons such as load, throughput, or efficient resource utilization. Alternatively, the application itself may monitor its performance and, when needed, ask the cloud to add or remove instances.
Nevertheless, we assume that the ISV defines a deployment policy that includes an upper bound to the number of application enclaves that can run simultaneously.~\footnote{The deployment policy may also define other constraints, e.g., number of enclaves running during day/night time, etc.} This is needed to mitigate forking attacks and \our must ensure that the ISV deployment policy is fulfilled at all times.

\noindent\textbf{Storage.} \our leverages a Byzantine fault tolerant storage instantiation based on MinBFT~\cite{MinBFT}. We opt to rely on MinBFT owing to its small code base. We assume a Key-Value storage abstraction~\cite{Lam86} which exports two operations: \textsc{PUT}($k,v$), which stores value $v$ indexed by key $k$, and \textsc{GET}($k$), which returns the stored value indexed by key $k$. We assume that the default value for any key is a special value, which is not a valid value for a PUT operation. We also assume that \textsc{PUT} and \textsc{GET} operations can only be invoked by authorized clients.

BFT storage is primarily used to prevent forking attack against \our. Nevertheless, applications can also leverage the storage service to keep either immutable state (e.g., the private key of a TLS server~\cite{talos}), and/or mutable state (e.g., a key-value store~\cite{brenner16middleware} that is read/written by all the application enclaves throughout their lifecycle). Indeed, secure storage offered by SGX (i.e., sealing) only allows for local storage and if several enclave applications require access to common storage, this must be provided as an additional service.

\subsection{Threat Model}\label{sec:threat}

The goal of the adversary that we consider is two-fold. On the one hand, the adversary may abuse the enclave provisioning process of \our in order to leak application secrets. On the other hand, the adversary may be interested in deploying a large number of application enclaves (i.e., larger than what is allowed by  that application's ISV) in order to amplify the effect of a forking attack against a victim application.

The adversary can compromise privileged code on a node and we denote that node as \emph{compromised}. However, we include SGX in the TCB and therefore assume that the adversary cannot compromise SGX components (e.g., system or application enclaves) on the compromised node.

We allow the adversary to compromise any number of nodes that host application enclaves or cloud management services. However, we only allow the adversary to compromise up to $f$ out of $2f+1$ nodes of the BFT storage layer. We argue that assuming a threshold to the storage nodes that an adversary can compromise is reasonable since compromising storage nodes (where no client-code can be deployed) is sensibly harder than compromising nodes where (malicious) clients can deploy their code. This assumption is in line with previous work on distributed BFT systems~\cite{DGLV10,ACKM06,Phalanx} and with previous work on forking attacks against TEEs~\cite{matetic17sec}. Further, this assumption is unavoidable since no secure distributed storage is feasible when all storage nodes are compromised. Even if one would na\"{i}vely fit the entire logic of a storage node in an enclave, realizing dependable storage would still require the assumption that at least one of the storage nodes is not compromised. Nevertheless, fitting the entire logic of a storage node in an enclave, leads to a large attack surface thereby weakening the assumption that enclave code is not susceptible of compromise. Splitting the logic between enclave and non-enclave code is the choice of all BFT protocols that leverage TEEs~\cite{DGLV10,ACKM06,Phalanx}.

We also assume that the adversary controls the network and as such controls the scheduling of all transmitted messages. Finally, we do not consider DoS attacks and we do not take into account attacks specific to SGX, such as the ones that exploit side-channels~\cite{brasser17woot}. We note that measures to mitigate attacks against SGX are orthogonal to \our\ and could be deployed alongside our solution.

\subsection{Overview}
\label{sec:design_and_overview}
To the best of our knowledge, there is no mechanism that enables enclave replication in a way that is transparent to the enclave owner. Clearly, a cloud provider can autonomously start an arbitrary number of enclaves as long as they do not require any secret material, nor do they need to access any confidential state information. However, as soon as an enclave requires a secret key (e.g., a TLS server such as Talos~\cite{talos}) or access to some confidential state (e.g., an encrypted key-value store such as SecureKeeper~\cite{brenner16middleware}), the enclave owner must be involved in the enclave startup process for attestation and secret provisioning.

Apart from the functional requirement of an online application owner, automatic enclave deployment in the cloud faces a number of security challenges. Namely, if deployment of application enclaves is mainly handled by the cloud, an adversary that manages to compromise the cloud infrastructure may try to run multiple enclaves of a given application, in order to mount~\emph{forking attacks}~\cite{brandenburger17dsn}. The enclave replication service must, therefore, be constantly aware of the number and status of deployed enclaves for a given application.

If we aim at designing an enclave provisioning service that removes the burden of being constantly online from the application owner, we should ensure that such service warrants its correct behavior to application owners and that confidentiality of the secrets is maintained in all the steps of the provisioning chain: from the application owner, until the target application. The security provisions of SGX make such a platform a promising candidate for the service we aim to design. If the provisioning service runs in an enclave, application owners can attest its code to ensure that the secrets of their applications will be handled properly. After attestation, an application owner can securely upload the secret key of its application and its $\mathrm{MRENCLAVE}$ to the provisioning service. From this moment on, the provisioning service acts on behalf of the application owner, by attesting enclaves of that application, ensuring that their untampered code runs in an enclave on an SGX-enabled platform, and by provisioning the application secrets. The provisioning service must also make sure that enclaves are deployed according to a policy set by the application owner, in order to mitigate forking attacks.

In our design, the provisioning service ensures that the ISV deployment policy is respected, but it does not decide \emph{when} an enclave for a given application should be provisioned or decommissioned. The provisioning service should only \emph{assist} the cloud when provisioning or decommissioning takes place. Namely, the decision to add or remove enclaves may involve business logic specific to the cloud provider. We separate our provisioning service from any business logic, so that the same service code may be used by several  cloud providers. Furthermore, our design facilitates the use of open-source code that can be audited via remote attestation or publicly vetted.

We augment the cloud software stack with a layer named Enclave Management Layer (\layer), dedicated to elastic enclave provisioning. \layer is in charge of provisioning and decommissioning enclaves on behalf of application owners. \layer is designed to run entirely in SGX so that \emph{(i)} application owners can verify its code and \emph{(ii)} sensitive data entrusted by application owners to \layer is shielded by any other software running on the same host.

\begin{figure}[tb]
	\centering
	\includegraphics[width=0.9\linewidth]{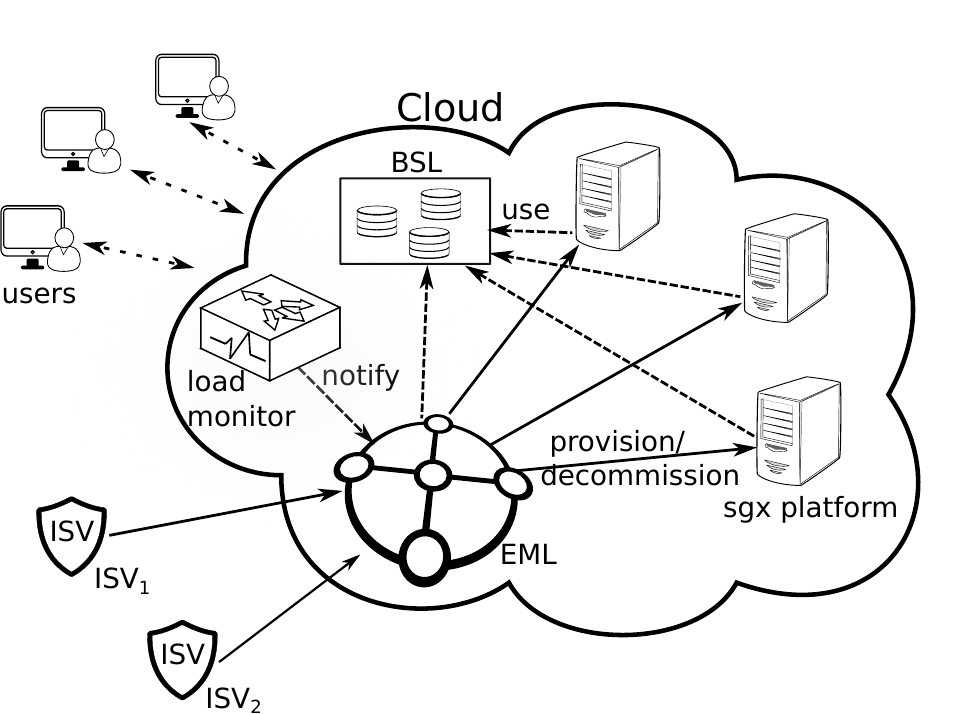}
	\caption{Sketch of \our system model. Independent Software Vendors (ISV) upload applications that may serve third-party users. The cloud monitors the load of applications and decides whether to add or remove enclaves for an application. This operation is carried out with the assistance of the Enclave Provisioning Layer (\layer). The latter leverages the Byzantine Fault-Tolerant Storage Layer (\storage) that can be also used by applications.}
	\label{fig:system}
\end{figure}

\layer is distributed across enclaves and leverages a master-slave approach to ensure progress despite potential crashes. Since \layer itself may be victim of forking attacks, we couple it with a BFT Storage Layer (\storage) that provides consistent storage despite Byzantine faults of a fraction of its nodes. \layer uses \storage to maintain at all times a consistent view of the requests to provision/remove enclaves and the progress it has made to handle such requests. This design allows us to prevent forking attacks on \layer while, at the same time, keeping the code-base of the provisioning service small enough to be run entirely in an enclave. Coupling a lightweight management layer such as \layer and a BFT storage layer such as \storage, we enable the cloud to dynamically provisions enclaves to applications, while ensuring protection against forking attacks.
Our solution is depicted in Figure~\ref{fig:system}. In a nutshell, application owners entrust the cloud provider with the application code, and \layer with the secret material that the application needs to run (e.g., a secret key).
When a new application enclave must be provisioned, \layer acts on behalf of the application owner and ensures that \emph{(i)} the deployment of the new enclave does not violate the policy defined by the application owner, \emph{(ii)} the application code runs in an enclave on an SGX-enabled platform, and that \emph{(iii)} the enclave is provisioned with the appropriate secret key, if required. When dealing with enclave decommissioning, we note that one cannot tell whether an enclave has been properly shut down or whether its messages are being blocked. To solve this issue, each application enclave is granted a lease upon provisioning. That is, when \layer provisions an application enclave, it also provides an ``end-of-lease'' timestamp. The application enclave should run until the lease expires, unless the lease is otherwise renewed.

\section{Protocol Specification}\label{sec:one}

Before describing \our in detail, we start by outlining the process of remote proxied attestation which constitutes an essential building block that will be used in our solution.

\subsection{Remote Attestation by Unregistered Verifiers}
\label{sec:proxy_attestation}
As mentioned in Section~\ref{sec:background}, the Intel Attestation Service (IAS) controls that remote attestation is not abused by verifiers and, in particular, that SGX platforms are not tracked---which constitutes one of the main goals of Direct Anonymous Attestation~\cite{brickell04ccs}. Nevertheless, involving IAS as an intermediary in each remote attestation limits the adoption of this mechanism, especially by parties who are not registered with IAS. This limitation becomes especially relevant if the enclave runs a public service like a mail server. Indeed, it is rather unrealistic to assume that all users interested in setting up a mail account are registered to IAS; yet, users may want to attest the code of the mail server and ensure it runs in an enclave on an SGX-enabled platform.

\begin{figure}[tb]
\centering
\includegraphics[width=\linewidth]{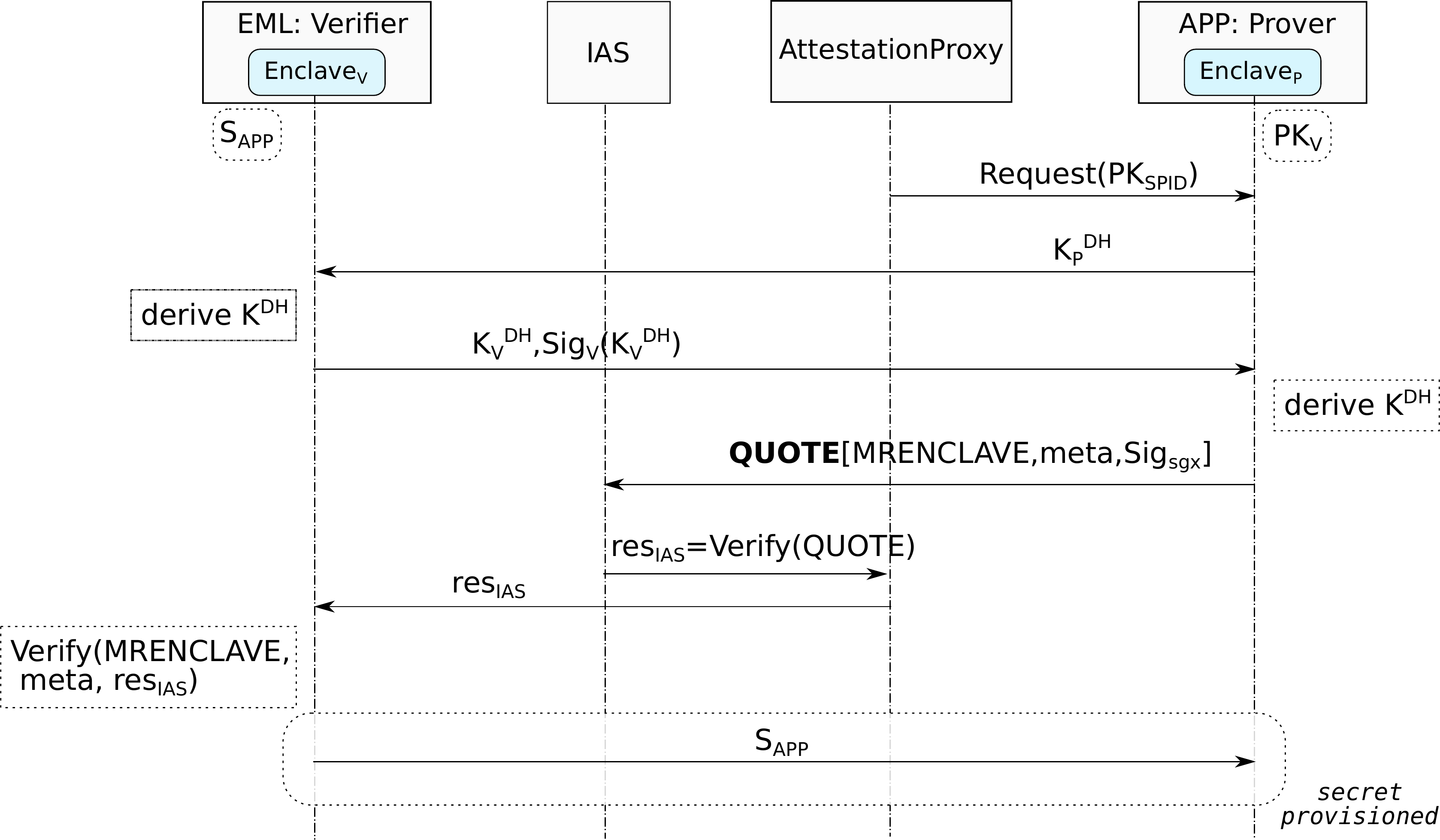}
\caption{Proxied attestation protocol. The DH ephemeral key of the verifier ($K_V^{DH}$) is signed either by the verifier itself (as shown in the figure) or by the proxy. In the former case, the enclave must have the public key of the verifier. If the ephemeral key is signed by the proxy, the enclave must have the public key of the proxy.\label{fig:proxy_attestation}}  %{\color{blue}Here, EPID refers to ``Enhanced Privacy ID'' which is used by IAS to prevent the association between the hardware and the quote, while SPID refers to the ID of the service provider that is registered to IAS.}
\end{figure}

In order to overcome this limitation and enable remote attestation with unregistered verifiers, we utilize a proxy registered to IAS. The proxy can be deployed by the cloud provider or by a third-party. Our proxied attestation protocol is depicted in Figure~\ref{fig:proxy_attestation}. There, we only provide an overview of the protocol; detailed message contents refer to the ones defined in the Intel SGX SDK Developer Reference~\cite{sgx-sdk}. Attestation via our proxy comes in two flavors, depending on whether the prover enclave ``knows'' (i.e., holds the public key of) the remote verifier. If the verifier is known to the prover, the proxy simply relays messages between prover and verifier; when the prover outputs and encrypted quote, the proxy (registered to IAS) forwards the ciphertext to IAS in order to get back the cleartext and provides the latter to the verifier. In case the verifier is unknown to the prover, the proxy also signs the ephemeral DH key chosen by the verifier. Therefore, the prover enclave must embed the public key of the proxy.~\footnote{Remote attestation using the standard SDK requires the ephemeral DH key of the prover to be signed and it also requires that the prover has the corresponding public key.}

Our proxied attestation protocol allows any party to remotely attest an enclave and to establish an unilaterally or mutually authenticated DH key---depending on whether the identity of the verifier is known to the prover.

Note that our protocol is compliant with the standard attestation protocol that leverages the SDK provided by Intel for SGX and only require the enclave developer to include the public key of the proxy, in cases where attestation requests are expected from unknown verifiers.

\subsection{\our Protocol Details}
\label{sec:details}
%\subsubsection{\layer}
\noindent\textbf{Setup.} Recall that \our comprises two layers: a BFT storage layer named \storage and an enclave provisioning layer called \layer. We assume that
\storage is setup initially by the provider conforming with the setup of MinBFT~\cite{MinBFT}. The setup of \layer unfolds as follows. The cloud provider \cloud (or a third party) starts $N$ enclaves, each running an instance of \layer. The enclaves must attest each other and agree on a group key for secure group communication. For this task, we require each \layer enclave to be aware of the identity of its peers (in order to attest them) and of the number $N$ of enclaves that form \layer. We use the group key exchange protocol by~\cite{Burmester05} and denote by $k_\layer$ the established group key. Note that attestation rules out active attacks. That is, SGX attestation ensures that only an instance of \layer enclave running in an SGX environment can participate in the key agreement protocol. Once \layer has been set up, the enclaves jointly generate a key-pair for a signature scheme and publish the verification key. Application owners must embed this key in their applications, in order to enable application enclaves to verify the legitimacy of the messages received from \layer during attestation.

\layer enclaves are organized in a Master-Slave approach. By default, the master enclave is the enclave that has the largest enclave identifier. During normal operation, the master is in charge of carrying out the main operations in \layer while slaves simply assume a passive role.

\layer's master implements a variant of the so-called ``node guarding'' protocol to keep track of the availability of the slaves; this essentially consists of the master exchanging alive messages with the slaves at regular intervals. The master enclave periodically sends a beacon request to the slaves to transmit information about their current state (e.g., stopped, active). If a slave does not respond to the request of the master within a certain timeout, the master considers the slave to be crashed and relays this information to the remaining slaves. On the other hand, the slaves also use this protocol to monitor availability of the master; if a request from the master is not received after a certain timeout, the slaves assume that the master itself has failed. In this case, the slave with the highest identifier from the set of active slaves, assumes the role of a master and starts issuing the beacon requests. This process ensures a continuous operation of \layer in spite of potential crash failures. Needless to mention, this entire lightweight protocol runs within the enclaves of \layer nodes. Further, if one of the \layer nodes crashes, it can be restarted and it can recover its state (e.g., $k_\layer$, \layer node endpoints, etc.) from the \storage storage layer.

\noindent\textbf{Notation.} We denote an application and its binary by $\alpha$ and $b_\alpha$, respectively. Also, $p_\alpha$ denotes the deployment policy defined by the application owner. In this paper, we assume the owner simply sets an upper bound to the number of enclaves that can run simultaneously. However, \our can be easily extended to account for more complex deployment policies.

\layer assigns identifiers to applications and enclaves. An identifier for an enclave of application $\alpha$ looks like $eid=\alpha||mr_\alpha||h_\alpha$, where $mr_\alpha$ is the $\mathrm{MRENCLAVE}$ of the application, and $h_\alpha$ corresponds to the hash of the key established between $\layer$ and the enclave during attestation.

In order to keep track of applications and enclaves, \layer leverages the storage functionality offered by \storage. In particular, for each application $\alpha$, \layer keeps track of the following metadata:
\begin{enumerate}
  \item $p_\alpha$: Upper bound to number of running enclaves.
  \item $mr_\alpha$: $\mathrm{MRENCLAVE}$.
  \item $ak_\alpha$:  Application secret key.
  \item $enc_\alpha$: A list of tuples ($eid, key, st, eol$) where $eid$ is an enclave identifier, $key$ is the key established between \layer and the enclave during attestation, $st$ is a status variable, and $eol$ is the current end-of-lease timestamp for that enclave. Variable $st$ can take values in $\{$att, run, tbd, tbs, sus$\}$. An enclave has status ``attested'' (att) after being attested by \layer. The status is changed to ``running'' (run) when the enclave is provisioned with the application secret key. The enclave status is set to ``to be deleted'' (tbd) or ``to be suspended'' (tbs) when the cloud requests the enclave to be deleted or suspended, respectively. Finally the status is set to ``suspended'' (sus) when the enclave has been suspended.
\end{enumerate}

\layer exports the identifiers of applications and enclaves to the cloud \cloud in order to efficiently manage enclaves for a given application. Note that application and enclave identifiers do not bear any sensitive information apart from the number of enclaves running for a given application---an information already available to the cloud.

We assume that the integrity and the confidentiality of data written/read to \storage (via PUT/GET) are protected by means of an authenticated encryption scheme. The key material for authenticated encryption is derived from the key shared by all \layer enclaves, namely $k_\layer$, by means of a key-derivation function.

We use application identifiers as the keys to the storage service and for each application we store a ``flat'' database to keep information of that application and its enclaves.
In order to ease exposition, we slightly overload the PUT/GET interface as follows.
We write \storageget{$\alpha;\ attr$} to fetch only the value of attribute $attr$ for application $\alpha$. Similarly, \storageput{$\alpha;\ attr\ :=\ value$} sets $attr$ to $value$, leaving all other attributes at the same key unchanged.
We also write \storageget{$\alpha;\ enc_\alpha : eid$} to fetch only the enclave information related to $eid$ (i.e., $key,\ st,\ eol$) from the list $enc_\alpha$. Also, \storageput{$\alpha;\ enc_\alpha : \langle eid',\ key',\ st',\ eol'\rangle$} writes to the list of enclaves $enc_\alpha$ of application $\alpha$; if $enc_\alpha$ already has a tuple with $eid==eid'$, this operation only updates the remaining fields to $key'$, $st'$, and $eol'$, respectively. If $enc_\alpha$ has no tuple with $eid==eid'$, then a new tuple $\langle eid',\ key', state',\ eol'\rangle$ is appended. We stress that this notation is only to improve the readability of our pseudocode. In reality, we always read and write the whole data associated to a given key.

\begin{algorithm}[t]
  \caption{Attestation of the \layer Service and Initial Upload of Code
    \label{alg:deploy}}
  \begin{algorithmic}[1]
    %\Require{\layer service with enclave $e_{\layer}$ is running; the owner of application $\alpha$ knows the $MRENCLAVE$ of $e_{\layer}$, namely $mr_\layer$.}
    \State \textsc{[Application owner]}
    \Function{Attest-Upload}{}
    	\State $k\leftarrow \textsc{ProxiedAttestation}(e_\layer,\ mr_\layer)$
        \If{$k_{\alpha,\layer}==\perp$}
            \State \Return{-1}
        \EndIf
    	\State \textsc{Send} $\langle p_\alpha,\ mr_\alpha,\ k_\alpha \rangle$ to $e_\layer$
    	\State \textsc{Send} $\langle b_\alpha \rangle$ to \cloud
    \EndFunction
    \State \textsc{[\layer]}
    \Function{Initialize}{$\alpha,\ p_\alpha,\ mr_\alpha,\ k_\alpha$}
    	\State \textsc{Put}($\alpha;\ p_\alpha,\ mr_\alpha,\ k_\alpha,\ \perp $)
    \EndFunction
  \end{algorithmic}
\end{algorithm}

\noindent\textbf{Attestation of \layer Service and Initial Upload of Code.} Algorithm~\ref{alg:deploy} lists the main steps carried out when an application owner wants to upload his application to the cloud. Before the application owner can entrust the management of his application to \layer, he must verify the identity of the \layer enclave and establish a secure channel. This is captured by the function $\textsc{ProxiedAttestation}(e_\layer,\ mr_\layer)$ that takes as input the endpoint of the enclave to be attested and the expected $\mathrm{MRENCLAVE}$. The function returns the key established with the prover enclave, if attestation is successful; otherwise it signals an error by returning $\perp$.

Once the application owner has established a secure channel with \layer, he uploads $p_\alpha,\ mr_\alpha,\ k_\alpha$ to \layer and $b_\alpha$ to \cloud. \layer writes $p_\alpha,\ mr_\alpha,\ k_\alpha,\ \perp$ to \storage and sends an acknowledgement to the application owner. The cloud stores $ b_\alpha$ and also sends an acknowledgement message to the application owner.

From this moment on, the application owner goes offline, while \layer cooperates with \cloud in order to increase or decrease the number of enclaves allocated to that application. \cloud can, at any time, issue requests to \layer to deploy or remove an enclave. Similarly, \cloud can ask to suspend a running enclave or resume a previously suspended enclave. \layer writes requests to storage in order to serialize them. Then, \layer periodically reads from \storage in order to identify pending requests and dispatch them.

\begin{algorithm}[t]
  \caption{Deployment Request\label{alg:deployment}}
\begin{algorithmic}[1]
\Function{Provision\_Request}{$\alpha,\ e$}
    \State $mr_\alpha\leftarrow $ \storageget{$\alpha;\ mr_\alpha$}
    \State $k_{\alpha,\layer}\leftarrow \textsc{ProxiedAttestation}(e,\ mr_\alpha)$
    \If{$k_{\alpha,\layer}==\perp$}
        \State \Return{-1}
    \EndIf
    \State $h_\alpha\leftarrow H(k_{\alpha,\layer})$
    \State $eid \leftarrow \alpha || mr_\alpha || h_\alpha$
    \State \storageput{$\alpha;\ enc_\alpha : \langle eid,\ k_{\alpha,\layer},\ \text{att},\ \perp\rangle$}\label{alg:deployment:writereq}
    \State \textsc{Send}($\text{ack},\ \alpha,\ e,\ eid$) to \cloud
\EndFunction
\end{algorithmic}
\end{algorithm}

\noindent\textbf{Deployment Request.} At this stage, the cloud provider creates a new enclave $e$ on an SGX platform and loads the code $b_\alpha$. It then contacts the \layer enclave that is acting as master to trigger the attestation and provisioning of the enclave.
The pseudocode of the steps carried out is provided in Algorithm~\ref{alg:deployment}.
Upon receiving a request, \layer enclave attests the application enclave (line 3) and assigns it an identifier made of the application identifier, the enclave identity, and the hash of the key established with that enclave during attestation (line 8). Next, \layer enclave writes to storage tuple $\langle eid,\ k_{\alpha,\layer},\ \text{att},\ \perp\rangle$ to reflect the fact that enclave $eid$ was attested and it is ready for provisioning. Finally, \layer enclave acknowledges to \cloud the end of the operation. If \cloud does not receive an acknowledgement within a given timeout, then \cloud may infer that the \layer enclave handling the request has crashed and that the request should be issued to another \layer enclave.

\begin{algorithm}[t]
  \caption{Termination Request\label{alg:termination}}
\begin{algorithmic}[1]
\Function{Terminate\_Request}{$eid$}
    \State Parse $eid$ as $\alpha||mr_\alpha||h_\alpha$
    \State $\langle key,\ st,\ eol\rangle \leftarrow \storageget{\alpha;\ eid}$
    \If{$st==run$}
        \State \storageput{$\alpha;\ enc_\alpha : \langle eid,\ key, \text{tbd},\ eol\rangle$}\label{alg:termination:writereq}
    \EndIf
    \State \textsc{Send}($\text{ack},\ \text{tbd},\ eid$) to \cloud
\EndFunction
\end{algorithmic}
\end{algorithm}

\begin{algorithm}[t]
  \caption{Suspension Request\label{alg:suspension}}
\begin{algorithmic}[1]
\Function{Suspension\_Request}{$eid$}
    \State Parse $eid$ as $\alpha||mr_\alpha||h_\alpha$
    \State $\langle key,\ st,\ eol\rangle\leftarrow \storageget{\alpha;\ eid}$
    \If{$st==run$}
        \State \storageput{$\alpha;\  enc_\alpha : \langle eid,\ key,\ \text{tbs},\ eol\rangle$}\label{alg:suspension:writereq}
    \EndIf
    \State \textsc{Send}($\text{ack},\ \text{tbs},\ eid$) to \cloud
\EndFunction
\end{algorithmic}
\end{algorithm}

\begin{algorithm}[t]
  \caption{Resumption Request\label{alg:resumption}}
\begin{algorithmic}[1]
\Function{Resumption\_Request}{$eid$}
    \State Parse $eid$ as $\alpha||mr_\alpha||h_\alpha$
    \State $\langle key,\ st,\ eol\rangle\leftarrow \storageget{\alpha;\ eid}$
    \If{$st==sus$}
        \State \storageput{$\alpha;\  enc_\alpha : \langle eid,\ key,\ \text{tbr},\ eol\rangle$}\label{alg:resumption:writereq}
    \EndIf
    \State \textsc{Send}($\text{ack},\ \text{tbr},\ eid$) to \cloud
\EndFunction
\end{algorithmic}
\end{algorithm}

\noindent\textbf{Termination/Suspension/Resumption Requests.} The pseudocode to terminate, suspend or resume an enclave is provided in Algorithms~\ref{alg:termination}, \ref{alg:suspension} and \ref{alg:resumption}, respectively. Requests are invoked by \cloud providing the enclave identifier $eid$ as an argument. The \layer enclave handling the request extracts the application identifier from $eid$ and fetches from \storage attributes $key,\ st,\ eol$ of enclave $eid$. For enclave termination, the \layer enclave checks that $st$ is ``run'' and sets it to ``tbd'' (i.e., to be deleted). For enclave suspension, the \layer enclave checks that $st$ is ``run'' and sets it to ``tbs'' (i.e., to be suspended). For enclave resumption, the \layer enclave checks that $st$ is ``sus'' and sets it to $tbr$ (i.e., to be run).

For termination and suspension of an enclave, \layer only takes note of the request by setting the status variable of that specific enclave; the operation is actually completed at the beginning of the next lease. This is because, as we argued above, there is no guarantee that the cloud is effectively terminating or suspending the enclave at the time of the request. However, the enclave will stop working at the end of the current lease and its lease will not be renewed as shown below.

For enclave resumption, once again \layer persists the request to storage by setting the status variable of that specific enclave; the enclave will be resumed by the main routine of \layer that dispatches provisioning and resumption requests persisted to storage (see next).

\begin{algorithm}[t]
  \caption{Dispatch\label{alg:dispatch}}
\begin{algorithmic}[1]
\Function{Run}{$\alpha$}
    \State $\langle p_\alpha,\ mr_\alpha,\ k_\alpha,\ enc_\alpha\rangle \leftarrow $ \storageget{$\alpha$}
    \State $\langle eid,\ key,\ st,\ eol\rangle \leftarrow \textsc{FindNext}(enc_\alpha)$\label{findnext}
    \If{$eid!=\perp$}
        \If{$\textsc{CountRunning}(enc_\alpha)<p_\alpha$}\label{alg:dispatch:count}
            \If{st\ ==\ \text{att}}
                \State \textsc{Send}($k_\alpha,\ eol$) to $eid$\label{alg:dispatch:provision}
            \Else \Comment{$st\ ==\ \text{tbr}$}
                \State \textsc{Send}($\text{resume},\ eol$) to $eid$\label{alg:dispatch:resume}
            \State \storageput{$\alpha;\ \langle eid,\ key,\ \text{run},\ eol\rangle$}\label{alg:dispatch:done}
            \EndIf
            \State \textsc{Send}($\text{ack},\ \text{run},\ eid$) to \cloud
        \EndIf
    \EndIf
\EndFunction
\end{algorithmic}
\end{algorithm}

\noindent\textbf{Enclave Provisioning/Resuming.} The pseudocode to dispatch requests to provision or resume enclaves is shown in Algorithm~\ref{alg:dispatch}. This code is periodically executed by the \layer enclave acting as master. Function $\textsc{FindNext}(enc_\alpha)$ on line 3 takes as input the list of tuples storing information about the enclaves of application $\alpha$ and returns the first tuple $\langle eid,\ key,\ st,\ eol\rangle$ such that the status variable $st$ is either ``att'' or ``tbr''. Status ``att'' means that the enclave has been attested and it is ready to be provisioned with the application secret key. Status ``tbr'' reflects a suspended enclave that must be resumed.
Before dispatching the request for $eid$, the \layer enclave checks that the number of running enclaves is below the upper bound set by application owner and that provisiong/resuming $eid$ does not violate the owner's constraints. Counting is carried out by function $\textsc{CountRunning}(enc_\alpha)$ on line 5. An enclave is considered as running if its status variable is set to ``running, ``to be suspended'', or ``to be deleted''.
Next, \layer enclave either provisions $eid$ with the application secret key and the current end-of-lease timestamp, or it sends to $eid$ a ``resume'' directive with the current end-of-lease timestamp. Finally, \layer writes to \storage that the enclave has been served and notifies \cloud.

From this moment on, the application enclave runs as expected, e.g., executing computation on behalf of the application owner or serving requests from clients. However, we require the application to halt its execution if the current time has passed the current end-of-lease timestamp received by \layer. Recall that a secure source of time is currently available on all SGX platforms via the \texttt{sgx\_get\_trusted\_time()} API.

\begin{algorithm}[t]
  \caption{Lease Renewal\label{alg:renewal}}
\begin{algorithmic}[1]
\Function{Renew}{$\alpha$}
    \State $\langle p_\alpha,\ mr_\alpha,\ k_\alpha,\ enc_\alpha\rangle \leftarrow $ \storageget{$\alpha$}
    \For{$\langle eid,\ key,\ st,\ eol\rangle$ in $enc_\alpha$}
        \If{$st\ ==\ \text{tbs}$}
            \State \storageput{$\alpha;\ \langle eid,\ key,\ \text{sus},\ eol\rangle$}\label{alg:renewal:sus}
        \ElsIf{$st\ ==\ \text{tbd}$}
            \State $\textsc{Delete}(enc_\alpha,\ eid)$\label{alg:renewal:del}
        \ElsIf{$st\ ==\ \text{run}\ \&\&\ eol\ <\ eol'$}
                \State \textsc{Send}(\text{renew}, $eol'$) to $eid$\label{alg:renewal:renew}
                \State \storageput{$\alpha;\ enc_\alpha : \langle eid,\ key,\ \text{run},\ eol'\rangle$}
        \EndIf
    \EndFor
\EndFunction
\end{algorithmic}
\end{algorithm}

\noindent\textbf{Lease Renewal.} The pseudocode shown in Algorithm~\ref{alg:renewal} is run by the \layer enclave acting as master when the current end-of-lease timestamp is approaching. At this stage, the \layer enclave scans through the list of enclaves belonging to application $\alpha$ and checks their status in order to determine whether the application must be suspended (line 4-5), deleted (lines 6-7), or whether its lease must be renewed. In the latter case, the application enclave receives the new end-of-lease timestamp $eol'$ with a ``renew'' directive. Regardless of the operation, the \layer enclave pushes the changes to \storage in order to persist the fact that the request was handled. Note that function $\textsc{Delete}(enc_\alpha,\ eid)$ on line 7 removes from $enc_\alpha$ the tuple referring to $eid$ and writes the updated list of tuples to storage.

\subsection{Dealing with Application Shared State}
\label{sec:shared}
Recall that some applications need to keep state to ensure its correct operation. Indeed, in a model where the cloud runs applications that span several enclaves, a shared storage layer might be required. This is because the sealing functionality of SGX is designed only to keep \emph{local} state and does not allow state to be shared across enclaves. In this case, newly provisioned enclaves should maintain a consistent view of such a state---otherwise the security of the overall service might be at risk. For example, in S-NFV~\cite{shih16sdnnfv}, the adversary could run two separated instances of the application and route state updates only to one instance, while exclusively pushing traffic flows to either instances. Hence, the outcome of processing a given flow may be different and dependant on whether it is carried out by one instance or the other. Similarly, password-strengthening services like Safekeeper~\cite{krawiecka18www} rely on rate-limiting to keep passwords secret. Having access to multiple isolated application instances, allows the adversary to infringe the restriction imposed by the rate-limiting policy.

\our's \storage can be used by such applications to share consistent state among their enclaves. Namely, whenever needed, authorized applications in \our can read/write their latest state from/to the storage layer using the offered PUT/GET interface. That is, our storage layer acts as consistent storage medium for various application enclaves to synchronize on their latest application state. For example, an enclave providing password strengthening service can continuously write the number of trials attempted on the storage layer. This allows to enforce rate-limiting across all application enclaves running the same service. In Section~\ref{sec:impl} we complement the evaluation of \our by assessing the overhead of using a BFT storage layer for applications that span across several enclaves.

\subsection{Security Analysis}
As mentioned in Section~\ref{sec:threat}, \our assumes an adversary that can compromise up to $f$ storage nodes and all nodes that run the applications. However, the adversary cannot compromise application enclaves since SGX ensures unhampered execution of code within enclaves as well as confidentiality of enclave data.

We note that even if the adversary compromises up to $f$ storage nodes, it cannot impact consensus realization in the storage layer. Namely, MinBFT~\cite{MinBFT} tolerates up to $f$ Byzantine nodes and ensures:
\begin{itemize}
\item Safety: all non-faulty storage nodes execute the requests in the same order (i.e., realize consensus).
\item Liveness: all clients (i.e., \layer enclaves) eventually receive replies to their requests.
\end{itemize}

In order to thwart forking attacks, one must ensure that \emph{(i)} \layer's log of the enclaves belonging to an application, namely $enc_\alpha$, reflects at all times the number and status of application enclaves deployed on \cloud, and \emph{(ii)} that the number of ``running'' enclaves in $enc_\alpha$ is compliant with the policy defined by the application owner. Naturally, we must also cater for the confidentiality of the application secret key throughout the application life-cycle.

\subsubsection{Secure Enclave Provisioning}

We start by analyzing the security of the enclave provisioning process in \our. Recall that this process allows the owner of an application $\alpha$ to securely provision his application enclaves by using the \layer service. In line with current SGX deployment models, the only piece of information that we regard as sensitive is the application secret key $ak_\alpha$, whereas the application binary is treated as non-sensitive data.

Before transferring the secret key of his application to \layer, the application owner must attest the \layer enclave and establish a secret key. This is done by leveraging the proxied attestation protocol of Section~\ref{sec:proxy_attestation}.
Note that during key establishment, the \layer enclave cannot attest the application owner (since the two parties may have not had any previous interaction). Therefore, \layer accepts application metadata (i.e., the application secret key, the policy, etc.) from \emph{any} party. Nevertheless we assume \cloud to authenticate application owners and that only authenticated application owners can contact \layer. This is a reasonable assumption since \cloud must authenticate application owners in order to bill them.

Once the application owner has securely uploaded the secret application key $ak_\alpha$ to \layer, the security provisions of SGX guarantee the confidentiality of the key while it is stored in the memory of the \layer enclave. If written to storage, $ak_\alpha$ is encrypted and authenticated with keys that are only available to \layer. Finally, \layer securely delivers the key to the application after attesting the code of the enclave and after establishing a secure channel.
Here again, attestation uses the proxied protocol of Section~\ref{sec:proxy_attestation}. However, attestation between \layer and an application enclave allows the enclave to authenticate \layer. This is achieved by embedding \layer's public key in the application code. By authenticating the prover, the application enclaves only accepts provisioning from \layer.

\subsubsection{Ensuring Consistency of \layer Operations}

We now analyze how \our ensures that the number of running enclaves, for a given application, is always below the bound set by the application owner through the deployment policy.
We achieve this by ensuring that all updates that affect the state of application enclaves are always registered by the storage layer that forms the backbone of \our.
Given that the storage layer implements a consistent Byzantine fault tolerant storage service, all registered events are totally ordered on consistent storage.
This design tolerates possible asynchrony or network partitioning that could arise in the \layer layer. Namely, since \layer enclaves do not run a consistent protocol (they only execute a lightweight node guarding protocol), consistency is guaranteed by the facts that \emph{(i)} all operations handled by \layer enclaves are duly registered on a consistent storage layer, and \emph{(ii)} all operations executed by \layer enclaves can be concurrently executed without the need for direct synchronization since the back-to-back execution of the same operation does not breach the security of \our. In what follows, we explain this in greater detail.

\noindent \textbf{Provisioning/Resuming:} Provisioning of an enclave $eid$ is only executed after the enclave has been attested and the request to be provisioned has been registered by writing the tuple $\langle eid,\ key,\ \text{att},\ \perp\rangle$ to \storage (Algorithm~\ref{alg:deployment}, line~\ref{alg:deployment:writereq}). This tuple reflects the fact that $eid$ has been attested and it is ready for provisioning.
Similarly, resuming of enclave $eid$ only occurs after the request has been registered by writing the tuple $\langle eid,\ key,\ \text{tbr},\ eol\rangle$ to \storage (Algorithm~\ref{alg:resumption}, line~\ref{alg:resumption:writereq}). Notice that in both cases, the tuple written to storage carries the key established with the $eid$ at the time of attestation. This allows any \layer enclave to establish a secure channel with that enclave in order to securely carry out requests.

Provisioning or resuming is carried out by Algorithm~\ref{alg:dispatch}. Since \storage ensures that write/read operations are serialized, no other enclave will be provisioned or resumed before the request for $eid$ is dispatched. This holds despite the fact that the \layer enclave in charge of handling the request for $eid$, say $e_\layer$, may fail, and despite the fact that multiple \layer enclaves may concurrently act as masters.

If $eid$ is to be provisioned and $e_\layer$ fails right after provisioning the application enclave (Algorithm~\ref{alg:dispatch}, line~\ref{alg:dispatch:provision}), the new master \layer enclave will use the same secret key $key$ to establish a secure channel with $eid$ and provision the application secret key once again.
Similarly, if $eid$ is to be resumed and $e_\layer$ fails right after sending the ``resume'' command (Algorithm~\ref{alg:dispatch}, line~\ref{alg:dispatch:resume}), the new master \layer enclave will use the same secret key $key$ to establish a secure channel with $eid$ and send once again the ``resume'' command.
In the above scenarios, we stress that provisioning or resuming \emph{the same} enclave does not violate the security provisions of \our.

We point out that even if two (or more) \layer enclaves acting as masters take in charge the request at the same time, they will both provision (or resume) $eid$. Also, they will both write the tuple $eid,\ key,\ \text{run},\ eol$ (Algorithm~\ref{alg:dispatch}, line~\ref{alg:dispatch:done}) to \storage in order to reflect that the operation has been dispatched. Once again, provisioning/resuming \emph{the same} enclave and writing \emph{the same} tuple to storage does not bring \our to an inconsistent state.

Only after the enclave status is set to ``run'' in \storage, \layer enclaves will start provisioning/resuming another enclave. This ensures that provisioning/resuming of enclaves is carried out in strict sequential order and allows \layer enclaves to be always aware of the running application enclaves for a given application.

\noindent\textbf{Terminating/Suspending:} As discussed before, once \cloud issues a request to terminate or suspend an enclave $eid$, there is no guarantee that the enclave has been effectively deleted or suspended. This is due to the fact that any attempt from \layer to contact $eid$ may be dropped by  the adversary that controls the communication network. For this reason, we resort to leases and require application enclaves to stop as soon as the current lease expires, unless \layer renews it.

\layer therefore treats an enclave $eid$ as suspended and sets its status accordingly (Algorithm~\ref{alg:renewal}, line~\ref{alg:renewal:sus}) only at the end of the lease. At the time \layer receives the request to suspend $eid$, it simply writes the request to \storage by setting $eid$'s state to ``to be suspended'' (Algorithm~\ref{alg:suspension}, line~\ref{alg:suspension:writereq}). However, the enclave is considered as running until the end of the current lease.
A similar approach is taken for requests to delete an enclave $eid$. The request is written to \storage by setting $eid$'s status to ``to be deleted'' (Algorithm~\ref{alg:termination}, line~\ref{alg:termination:writereq}), however the enclave will be considered as running until the end of the current lease. At that time, the enclave metadata is deleted from storage (Algorithm~\ref{alg:renewal}, line~\ref{alg:renewal:del}).

Note that enclaves considered as running (i.e., the ones with status set to ``running'', ``to be suspended'', or ``to be deleted'') affect the decision of whether a request to provision/resume an enclave should be completed. That is, an enclave is provisioned/resumed only if the number of enclaves considered as running  is below the threshold set by the application owner (Algorithm~\ref{alg:dispatch:count}).

\noindent\textbf{Lease Renewal: }At the end of a lease, \layer proceeds to renew the lease to all application enclaves with status ``running''. If a \layer enclave crashes after renewing the lease to a given enclave $eid$, but before writing to \storage that the operation was completed (Algorithm~\ref{alg:renewal}, line~\ref{alg:renewal:renew}, then $eid$ will receive the same renewal message from another \layer enclave taking up the master role. Once again, repeating the lease renewal operation issuing the same end-of-lease timestamp to the same enclave does not constitute a security breach.

\section{Performance Analysis}\label{sec:impl}
\subsection{Implementation Setup}

We deployed the storage service of \our on five identical servers with SGX supports. Each server is equipped with Intel Xeon E3-1240 V5 (8 vCores @3.50GHz) and 32~GiB RAM. The \layer instances were deployed on a machine with Intel Core i5-6500 (4 Cores @3.20GHz) and 8~GiB RAM. All these machines are equipped with SGX to run enclaves and are connected
 with a 1Gbps switch in a private LAN network. We argue that this setting emulates a realistic cloud deployment scenario where the compute servers and their corresponding storage servers communicate over the cloud's private LAN (e.g., Amazon AWS and S3).

As mentioned earlier, we instantiate the atomic storage service of \our using MinBFT. Our implementation of  MinBFT uses 2 interface functions ($\mathtt{createUI}$, $\mathtt{verifyUI}$~\cite{MinBFT}) and a total 339 LoC in SGX enclave in order to achieve Byzantine Fault Tolerance in the storage layer.\footnote{We contrast this to Paxos (based on LibPaxos [31]) which requires around 4,000 LoC.}
We argue that this is small enough to make formal verification of the consensus layer code base as needed. In our evaluation, we relied on HMAC-SHA256 MACs to achieve authentication between replicas and clients~\cite{pbft,MinBFT}. Notice that our evaluation only accounts for the normal case of MinBFT (i.e., we do not emulate Byzantine failures). In MinBFT, if $f+1$ replies correspond to a given version, then the version is committed. This masks transparently and by default up to $f$ failures. For these reasons, we stress that Byzantine failures do not affect performance of such classes of BFT algorithms.

We implemented the proxied attestation procedure described in Section~\ref{sec:proxy_attestation} based on the libraries provided by SGX SDK~\cite{sgx-sdk}. To establish a secure channel during provisioning, we rely on SGX's Diffie-Hellman key exchange library (256-bit ECC). In our proxied attestation implementation, the prover's code in the enclave requires around 200 lines, while the verifier's code in the \layer enclave is around 800 lines (cf. Table~\ref{tab:default}).\footnote{The verifier enclave also includes JSON and Base64 decoder libraries~\cite{json-decoder-lib, base64-decoder-lib} in order to decode the response from IAS.} In our implementation, we do not measure the latency incurred when communicating with the Intel Attestation service and we only measure the time of verifying the report issued by IAS.
\begin{table}[t]
\begin{center}
%\scalebox{0.8}{
\begin{tabular}{|c|c|}
\hline
\textbf{Application} & \textbf{Line of Codes (LoC)}\\
\hline
\hline
MinBFT & 339\\
Proxied attestation (prover) & 200\\
Proxied attestation \& provisioning (verifier) & 800\\
DupLESS integrated with \our & 80\\
\hline
\end{tabular}
%}
\end{center}
\caption{LoC required for implementing various routines of \our}\label{tab:default}
\end{table}

\begin{figure*}[tb]
    \centering
    \subfigure[Throughput vs. latency for enclave provisioning when $f=1$.]
    {
        \label{fig:provisionTP1}
        \includegraphics*[width=0.45\linewidth]{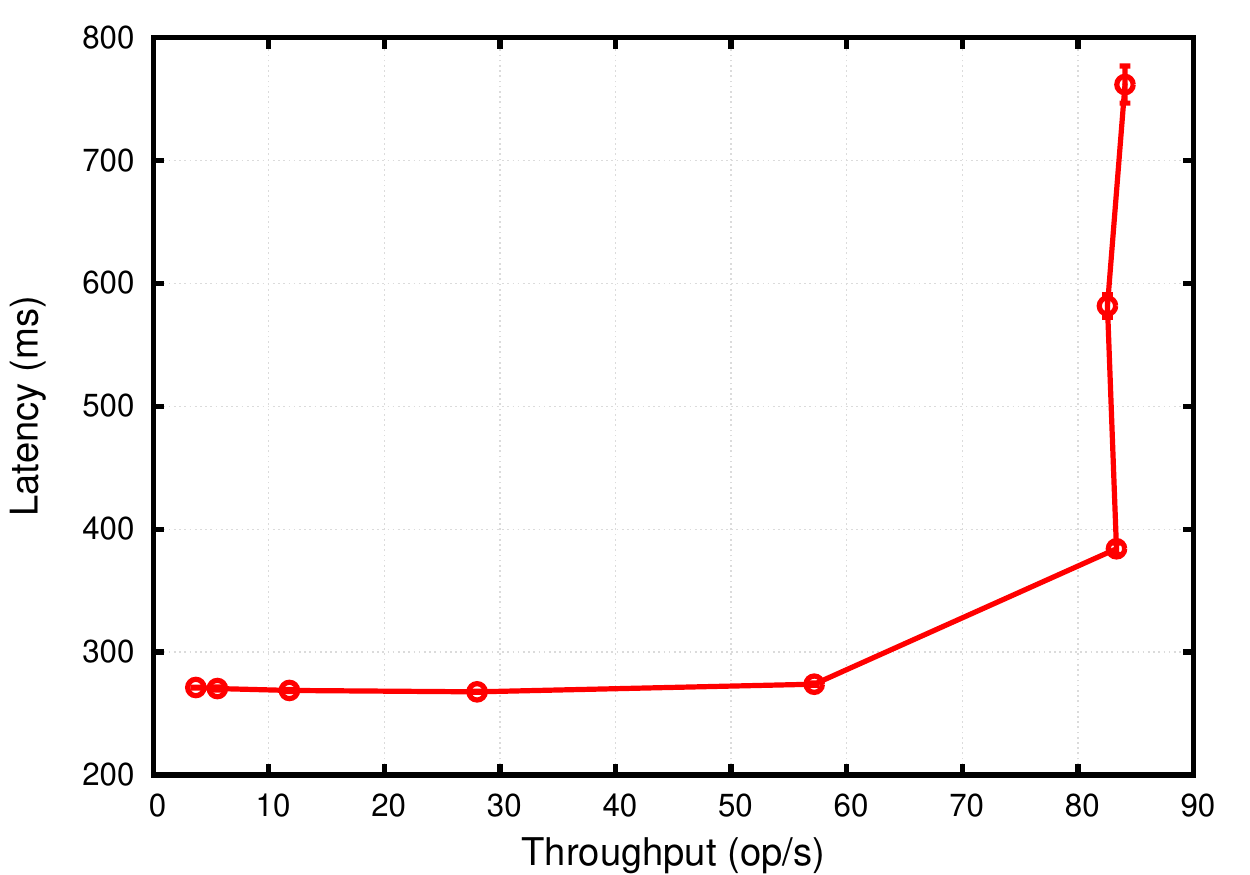}
    }
    \hspace{.01in}
    \subfigure[Throughput vs. latency for enclave provisioning when $f=2$.]
    {
        \label{fig:provisionTP}
        \includegraphics*[width=0.45\linewidth]{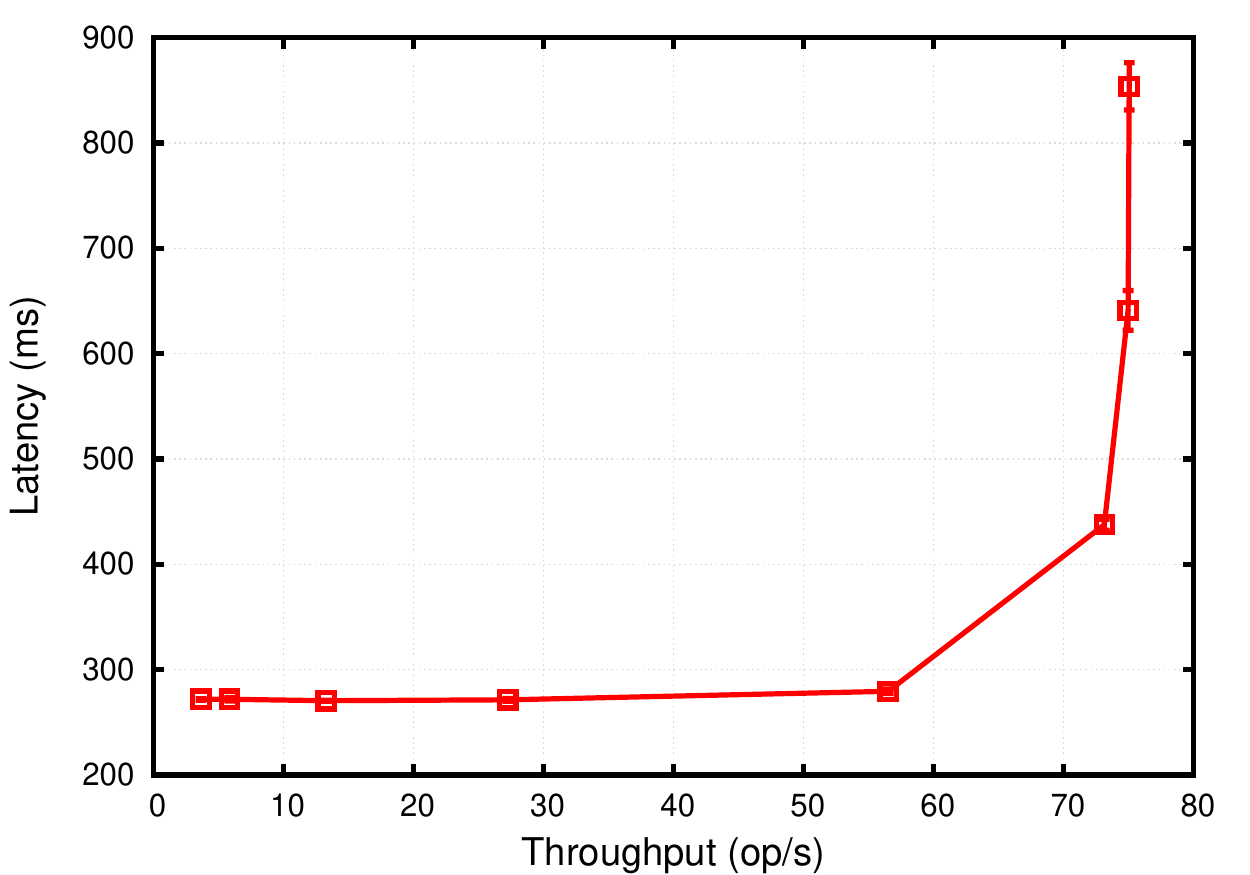}
    }
    \subfigure[Latency witnessed in the enclave provisioning process of \our.]
    {
        \label{fig:provision-repartition}
        \includegraphics*[width=0.45\linewidth]{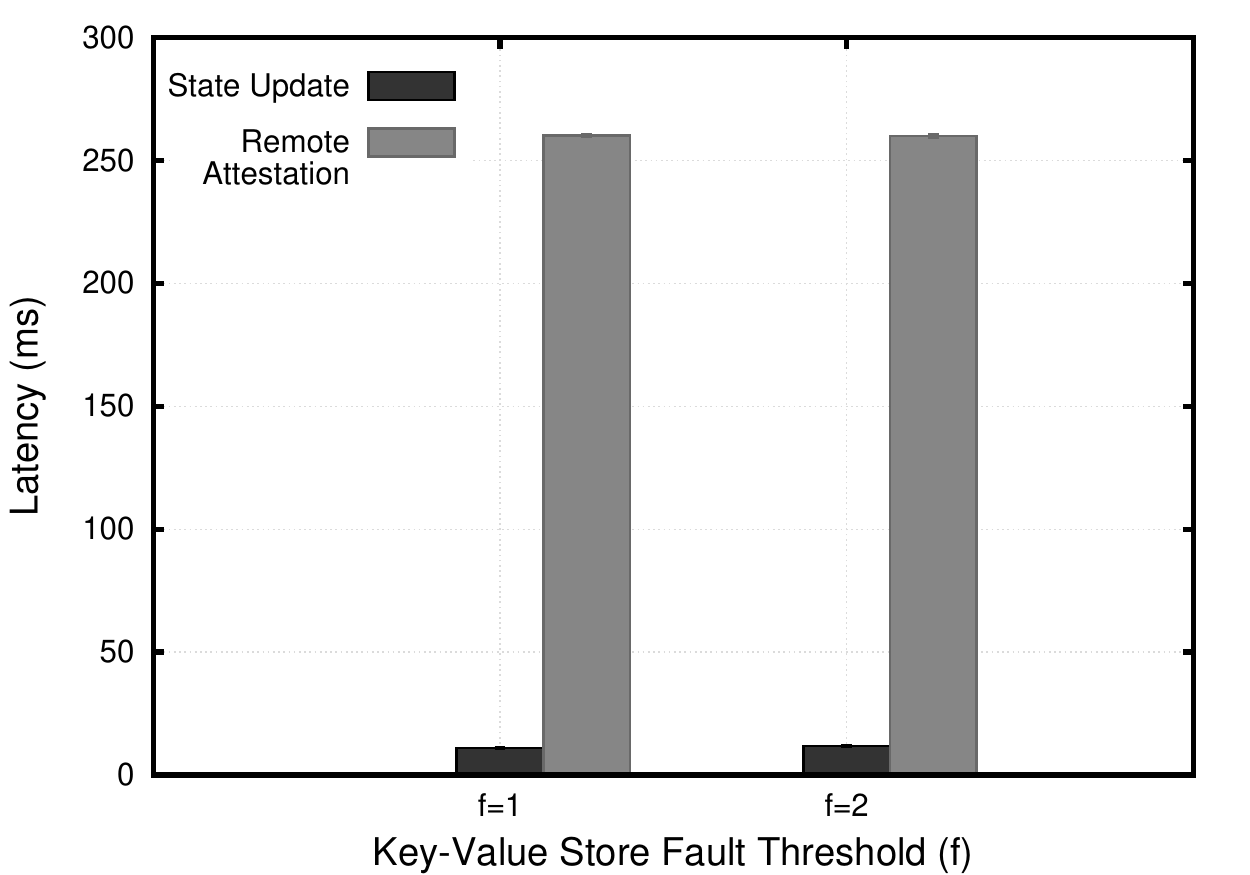}
    }
    \hspace{.01in}
    \subfigure[Throughput vs. latency for DupLESS w/ and w/o integration with \our.]
    {
       \label{fig:dupless}
       \includegraphics*[width=0.45\linewidth]{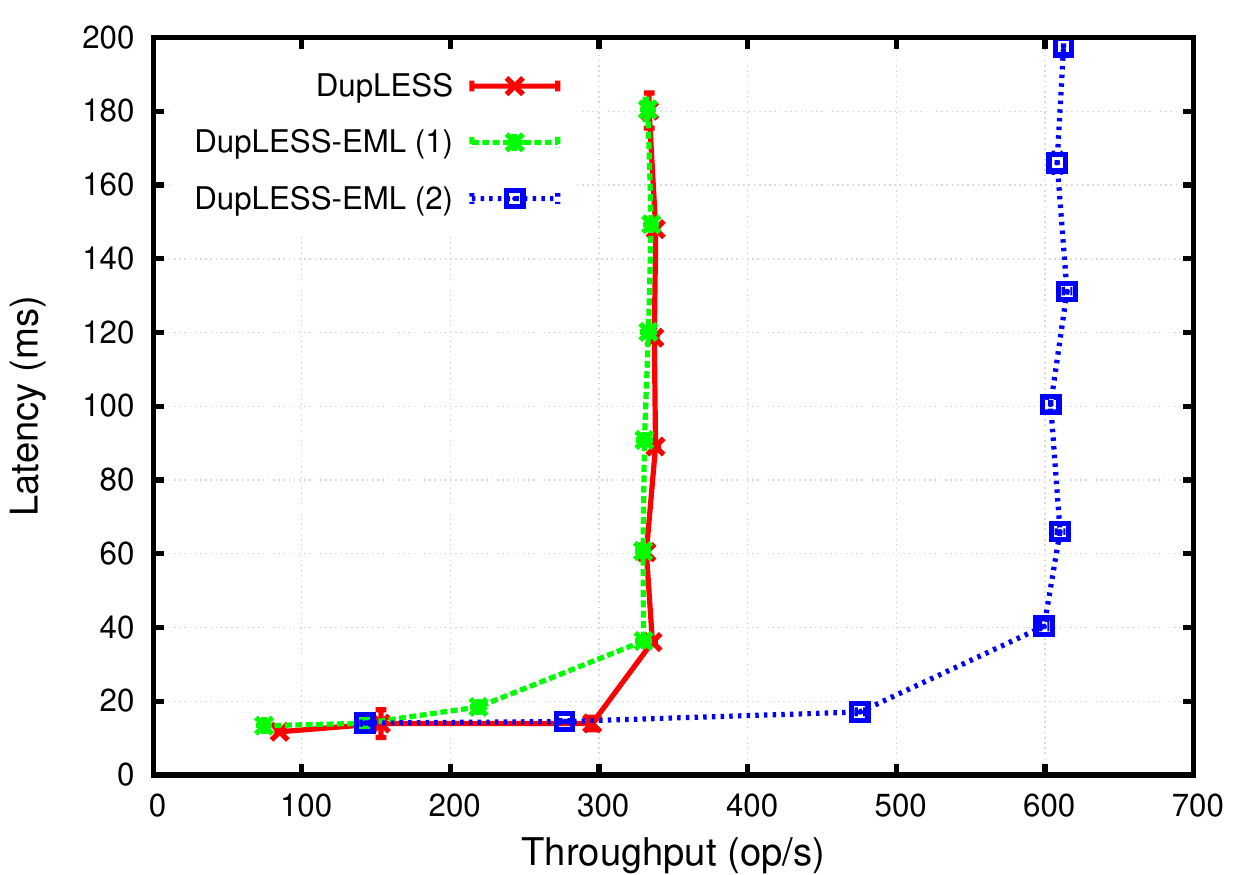}
    }
    \caption{Evaluation of the performance of \our in our setup. Data points are averaged over 10 independent runs; where appropriate, we include the corresponding 95\% confidence intervals.}
    \label{fig:eval}
\end{figure*}

\subsection{Evaluation Results}

In what follows, we evaluate the performance of \our in our setup. Namely, we measure the latency incurred in the provisioning of enclaves and in termination/suspension/resumption and lease renewal. Note that we do not evaluate the overhead incurred in the initial setup phase of \layer and the initial code upload by ISVs, since the setup is carried out only once and the overhead for ISVs to upload their code to the cloud is not particular to \our and is incurred by all applications that leverage cloud-based SGX deployments.

We also measure the latency incurred in the provisioning of enclaves with respect to the achieved throughput. We measure the throughput as follows. The master \layer enclave invokes operation in a closed loop, i.e., enclaves may have at most one pending operation. We require that the master \layer enclave performs a series of back-to-back operations (requests) and measure the end-to-end time taken by each operation. We then increase the number of provisioning requests in the system until the aggregated throughput attained by all requests is saturated.

%
%Since the cloud requests for instance management (e.g., provision/terminate/suspend/resume) is rather adhoc, we only measure the performance of enclave provisioning operation as the process of remote attestation is most expensive, while the other operations only incur state update in the key-value store.
%We evaluate the latency and throughput incurred in the enclave provisioning process (see Algorithm~\ref{alg:provision}) with respect to different server failure threshold $t$ of the key-value store. Similarly, we also evaluate the performance of DupLESS signing operation with and without synchronization with the key-value store.
%The throughput is defined as the number of operations that the system processes during a specific time interval.

\noindent\textbf{Enclave Provisioning:} In Figures~\ref{fig:provisionTP1} and~\ref{fig:provisionTP}, we evaluate the throughput vs latency for the enclave provisioning process given different storage failure threshold $f$. We see that when $f=1$ (3 storage servers), the system achieves a peak throughput of $85$~op/s with a latency of $270$~ms. On the other hand, when $f=2$ (5 storage servers), the latency remains almost the same, while the peak throughput is reduces to $75$~op/s. Our findings suggest that the remote attestation process is the dominant factor in the operation latency. Notice that even if increasing the fault-tolerance threshold of \storage reduces the peak throughput (since it requires more communication rounds), it has limited impact on the witnessed latency.

In Figure~\ref{fig:provision-repartition}, we further measure the constituent latencies incurred in the enclave provisioning process. In both cases when $f=1$ and $f=2$, we see that the time for remote attestation is around $260$~ms while the state update only takes $10$~ms without noticeable difference in either cases. Namely, the state update only comprises up to $3.7$\% of the whole provision process even when $f=2$.
%Considering that \layer still needs to request IAS to verify the attestation quotes, \wlnote{the time used in state update during the} enclave provisioning process is negligible. {\color{red}what?}

\noindent \textbf{Termination/Suspension/Resumption/Renewal Requests:} Recall that termination, suspension, resumption, and renewal requests basically consist of the \layer enclave updating the records corresponding to the target enclave on the storage layer. These requests are practically instantiated by a PUT request issued by the \layer primary enclave to update the associated record. In \our, such PUT requests only consume 0.86~ms with a peak throughput of $9800$~op/s when $f=1$ and 0.94~ms with a peak throughput of $4700$~op/s when $f=2$.
%{\color{red}this is inline with what we said before}

\noindent\textbf{DupLESS instantiation:} In Figure~\ref{fig:dupless}, we evaluate the performance overhead incurred by \our on applications that require shared mutable state for their correct operation. To this end, we implement a variant of DupLESS~\cite{dupless} and integrate it with \our in the case where $f=1$. DupLESS is a server-aided encryption scheme that enables data deduplication over encrypted data. In this scheme, users interested in deduplicating their files first contact the DupLESS gateway to obtain an encryption key that is derived to the file digest. This key is essentially a blind signature on the file digest that allows client to obtain encryption keys while keeping privacy of their files. By using a deterministic encryption scheme and a key derived from the file digest, two users with the same file will produce the same ciphertext that, as such, can be deduplicated by a storage service. By involving the gateway in the key generation process, brute-force attacks on predictable files can only be slowed down by rate-limiting the requests to the server. %In this sense, DupLESS implements a rate-limiting functionality which limits the number of requests issued by the same client over a period of time. More details on DupLESS can be found in Appendix~\ref{ap:dupless}.
In our variant implementation, we integrate DupLESS's blind signature scheme within SGX enclaves and use it as an exemplary application of \our.\footnote{We chose DupLESS because it incurs minimum I/O and allows us to clearly evaluate the computational overhead of \our.} Namely, we rely on \our to automatically commission and decommission DupLESS enclaves and to allow running enclaves to synchronize on their latest state to effectively enforce rate-limiting across all running enclaves. Since DupLESS leverages RSA-based blind signatures, we utilize the SGX-SSL library~\cite{sgx-ssl} to implement the signing functionality (with 4096-bit RSA) with \textasciitilde 80 lines of code. We deploy the DupLESS servers on a machine with Intel Xeon E3-1240 V5 and evaluate the overhead introduced by \our in this setting when compared to a standalone DupLESS gateway that does not leverage any functionality from SGX (i.e., the standard DupLESS gateway described in~\cite{dupless}).

Our results show that the latency incurred by a standalone DupLESS gateway is $18$~ms with a peak throughput of $330$~op/s. On the other hand, integrating a single DupLESS instance in \our achieves almost the same performance. This confirms that \our does not add significant overhead to existing SGX-based enclaves. Notice that adding an additional DupLESS enclave almost doubles the peak throughput by reaching around 600 op/s (for 2 DupLESS instances). The throughput exhibited by a distributed DupLESS instantiation will be however limited by the peak throughput exhibited by \storage which is roughly 9800 op/s; in this case, \storage can accommodate for roughly 30 DupLESS instances. We stress that replicating DupLESS using \our does not have any noticeable impact on the latency witnessed by DupLESS users.

%\footnote{Since the operation is computation intensive, when message rate surpasses the peak throughput, the waiting time becomes very large for every message and the system throughput drops dramatically to around $8,000$~op/s.}
%We see that since the operations in DupLESS mainly requires CPU resources, the system quickly saturates when there are more than $4$ concurrent requests. More specifically, the latency of the local operation is $0.26$ ms and the peak throughput $14,000$ op/s.
%We then evaluate the performance when the distributed storage layer is involved in each application request (e.g., to get a global view across difference instances for rate-limiting policy). We see that the latency increases to \wlnote{1.1} ms for $t=1$ and \wlnote{1.5} ms for $t=2$; similarly, the peak throughput reduces to \wlnote{7,800} op/s and \wlnote{4,500} op/s for $t=1$ and $t=2$ respectively. Notice that synchronization among the application instances reduces the system throughput almost by half since computation over the network is a costly operation.
%However, notice that in this experiment we did not apply optimizations such as batching the read/write requests to the storage consensus layer, which would reduce the communication cost and improve the throughput. For applications that relies on rate-limiting policies, batching optimization can be applied to certain extent in order to avoid restricting the upper-bound limit configured in the policy. {\color{red}please clarify how can batching be done}

\section{Related Work}\label{sec:related}
To the best of our knowledge, no previous study has addressed the problem of enabling seamless replication of SGX enclaves in the cloud. We now briefly review related work in the area.

Gu \emph{et al.}~\cite{gu17dsn} provide an SDK to enable enclave migration in the cloud. Here, enclaves are augmented with a thread that carries out state transfer. The thread in the source enclave brings other threads to a quiescent state and ships the internal state to the target enclave; a thread in the target enclave receives the state, installs it and recover execution. Since some state information is only available to the platform, the authors use a number of heuristics to estimate that part of the state and transfer it to the target platform. The authors show that their heuristic are indeed effective in few application scenarios. However, the effectiveness of this heuristic for general SGX applications remains to be assessed.

Matetic \emph{et al.}~\cite{matetic17sec} proposed a scheme, ROTE, to enable rollback protection for SGX enclaves. Recall that the sealing functionality of SGX provides confidentiality and integrity but does not guarantee freshness of sealed data. In a rollback attack, a malicious host leverages this shortcoming to provide enclaves with stale state information. In ROTE, a set of \emph{ROTE Enclaves} running on different platforms, help \emph{one} application enclave to maintain monotonic counters that, when used in conjunction with the sealing functionality of SGX, provide state freshness.
The set of ROTE enclaves is static and must be setup by an administrator before applications can leverage the service. Notice that ROTE does not deal with applications that span across several enclaves and requires that the application enclave runs on one of the platform that hosts ROTE enclaves.

ICE~\cite{Strackx14acsac} is another proposal that addresses rollback attacks in SGX. Differently from ROTE, ICE is a ``standalone'' solution that relies on hardware modifications to the platform, including dedicated on-chip registers backed by off-chip NVRAM.

Brandenburger \emph{et al.}~\cite{brandenburger17dsn} address forking attacks on TEEs in application scenarios where multiple clients interact with an enclave running at a malicious host. In order to counter forking attacks, they require an enclave to create a hash chain with the history of all performed operations. When combined with monotonic counters shared with all clients, such an approach can ensure fork linearizability~\cite{mazieres02podc}.

Proxied attestation was first proposed in~\cite{krawiecka18www}. Here, the proxy is registered with IAS and acts on behalf of the (unregistered) verifier towards the IAS. Notice that~\cite{krawiecka18www} leverages a \emph{proactive} attestation scheme where the enclave itself requests a quote from the platform and binds it to its ephemeral DH key \emph{before} seeing the ephemeral DH key of the verifier. This design saves round-trips during attestations but is not compliant with the SDK of Intel SGX; namely, a quote is provided \emph{after} the ephemeral DH key of the verifier has been received and a shared key established.\footnote{The data structure providing the quote is referred to as \texttt{msg3} in the SDK\cite{sgx-sdk} which is returned by \texttt{sgx\_ra\_proc\_msg2()} that processes the ephemeral DH key of the verifier and a valid signature on that ephemeral key.} Therefore, the scheme of~\cite{krawiecka18www} requires application developer to update their code in order to account for changes in the attestation protocol.
%The current SDK for SGX requires the ephemeral DH key of the verifier to be signed so that the prover enclave can authenticate the verifier. Since we could not access the details of the proxy proposed in~\cite{krawiecka18www}, we speculate that the proxy signs the DH key provided by the verifier, and that the proxy's verification key is available at the prover enclave in order to verify that signature.
Furthermore, the attestation protocol proposed in~\cite{krawiecka18www} only provides an unilaterally authenticated DH key exchange, since the enclave cannot be sure that the ephemeral DH key is the one chosen by the verifier and not by the proxy. Mutually authenticated DH key exchange would require the enclave to embed the verification key of the verifier. However, this is not viable if the enclave is meant to be verified by any (previously unseen) user of the cloud service.

%Regarding enclave migration, moving the internal state of an enclave to another platform is at odds with the main security provisions of SGX --- the state of an enclave is private to the enclave itself. Furthermore, when the state is persisted to disk, it can only be recovered by that same enclave running on that same platform.
%The only proposal for enclave migration that we are aware of, is by Gu et al.~\cite{gu17dsn}. It leverages an SDK that augments enclaves with a thread dedicated to migration. This thread simply transfers the internal state of the source enclave to the matching thread of the destination enclave. The authors of~\cite{gu17dsn} also point out that some data structures that must be migrated are not available to the enclave. This is the case of the CSSA -- a data structure that handles the nesting level of enclave exceptions. The solution proposed in~\cite{gu17dsn} is to infer this value by monitoring the behavior of the enclave and to rely on the untrusted OS to recreate the same conditions at the target platform.

\section{Conclusion}\label{sec:conclusion}

In this paper, we presented a novel solution, \our, that enables dynamic replication and de-commissioning of TEE-based applications in the cloud. \our leverages an SGX-based provisioning service that interfaces with Byzantine Fault Tolerant storage layer  to orchestrates dynamic application replication in the cloud without the active intervention of the application owner.

We showed that \our withstands a powerful adversary that can compromise a large fraction of the cloud infrastructure. By means of a prototype implementation, we also showed that \our moderately increments the TCB and does not add significant overhead to existing SGX-based applications.

\our therefore emerges as the first secure and practical solution to support elasticity of TEE-based applications in the cloud. As such, \our enables applications from benefitting from high availability, performance, and cost effectiveness that essentially form the basis of the cloud-computing paradigm.

\bibliographystyle{plain}
\bibliography{main}

\appendix
\section{MinBFT}\label{ap:minbft}

MinBFT comprises four routines and unfolds as follows:

\begin{enumerate}
   \item \textbf{Request}: Clients send their request messages asking the replicas to execute certain operations. A client $\mathcal{C}$ prepares its requested operation $op$ in message $\langle \mathsf{REQUEST}, \mathcal{C}, seq, op \rangle_{\sigma_{\mathcal{C}}}$, where $seq$ records the (local) message sequence from each client to prevent re-execution of the operations, and $\sigma_{\mathcal{C}}$ is the client signature.
   %a client $\mathcal{C}$ requests the execution of operation $op$ by sending to all replicas $m = \langle \mathsf{REQUEST}, \mathcal{C}, seq, op \rangle_{\sigma_{\mathcal{C}}}$, where $seq$ records the (local) message sequence from each client to prevent re-execution of the operations, and $\sigma_{\mathcal{C}}$ is the client signature. %that ensures the request is not forged.
   \item \textbf{Prepare}: This phase is triggered when the primary $\mathcal{S}_p$ receives a request message $m$. Once the request is validated, the primary asks its TEE to generate a unique message identifier $UI_p=\langle c, m \rangle_{\sigma_p}$. Note that the counter $c$ is monotonically increasing and the signature $\sigma_p$ is from the TEE. Subsequently, $\mathcal{S}_p$ multicasts $\langle \mathsf{PREPARE}, v, \mathcal{S}_p, m, UI_p, \rangle$ to the other replicas.
   %the primary $\mathcal{S}_p$ validates the request, and asks the secure hardware to generate a signed unique identifier $UI_p$ to the request message. Subsequently, $\mathcal{S}_p$ multicasts $\langle \mathsf{PREPARE}, v, \mathcal{S}_p, m, UI_p, \rangle$ to the other replicas, where $v$ is the current view \csnote{I am not really sure what ``view'' means.} that is maintained by the primary.
   \item \textbf{Commit}: This phase serves to acknowledge a valid $\mathsf{PREPARE}$ message. Each replica $\mathcal{S}_i$ responds with a $\mathsf{COMMIT}$ message. In particular, each replica multicasts $\langle \mathsf{COMMIT}, v, m, \mathcal{S}_i, UI_i, \mathcal{S}_p, UI_p \rangle$, where $UI_i$ is a unique identifier that $S_i$ gets from its TEE.
   \item \textbf{Reply}: A request is \emph{committed locally} and can be executed once a replica has received enough (i.e., $f+1$) consistent commits, because it is ensured that any request that commits locally on a correct replica will be committed on at least $f+1$ correct replicas eventually. Therefore, the replica can execute the operation $op$ and send the reply $\langle \mathsf{REPLY}, \mathcal{S}_i, seq, res \rangle$ with the execution result $res$ back to the client.
   %it is assured that at least each $\mathcal{S}_i$ executes the operation $op$ as soon as it receives $f+1$ commits on the same $m$ from different servers and sends the reply $\langle \mathsf{REPLY}, \mathcal{S}_i, seq, res \rangle$ with the execution result $res$ back to the client.
   \item \textbf{View-Change}: When a primary is suspected to be misbehaving, a replica can request a replacement of the primary through the view-change procedure. For example, when a received request failed to be executed within a certain timeout, a replica multicasts a view-change request $\langle \mathsf{REQ-VIEW-CHANGE}, \mathcal{S}_i, v, v' \rangle$, where $v'$ is the new view number and $v'=v+1$. If a replica receives $f+1$ $\mathsf{REQ-VIEW-CHANGE}$, it moves to view $v'$. At this stage the replica multicasts $\langle \mathsf{VIEW-CHANGE}, \mathcal{S}_i, v', CP, O, UI_i \rangle$, where $CP$ is the latest certificate and $O$ is the set of all messages sent by the replica since $CP$. Once the new primary of view $v'$ receives $f+1$ valid $\mathsf{VIEW-CHANGE}$ messages with consistent system state, the view change is executed by the new primary who broadcasts message  $\langle \mathsf{NEW-VIEW}, \mathcal{S}_{p'}, v', V_{vc}, s, UI_{p'} \rangle$, where $V_{vc}$ is the view-change certificate that includes all the received $\mathsf{VIEW-CHANGE}$ messages, and $s$ is the current system state which will serve as the initial state of view $v'$.
\end{enumerate}

The correctness of MinBFT holds as long as there is at least one honest node involved in any two quorums, thus only $2f+1$ replicas are required to tolerate $f$ faulty nodes. Further details on MinBFT can be found in~\cite{MinBFT}.

\section{DupLESS}\label{ap:dupless}

\begin{figure}[t]
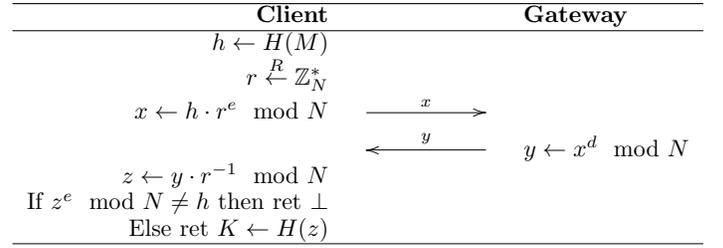

\begin{center}
\scalebox{0.9}{\begin{tabular}{rcl}
\hline
\textbf{Client} & \hspace{2cm} & \textbf{Gateway} \\
\hline
$h \leftarrow H(M)$ & &\\
$r \overset{R}{\leftarrow} \mathbb{Z}_N^*$ & &\\
$x \leftarrow h \cdot r^e \mod{N}$ & $\verylongrightarrowRF{x}$ &\\
 & $\verylongleftarrowRF{y}$ & $y\leftarrow x^d \mod{N}$ \\
$z \leftarrow y \cdot r^{-1} \mod{N}$ & &\\
If $z^e \mod{N} \neq h$ then ret $\perp$ & &\\
Else ret $K \leftarrow H(z)$ & &\\
\hline
\end{tabular}}
\end{center}
\caption{RSA blind-signature scheme adapted from~\cite{dupless}. $H: \{0,1\}^* \rightarrow \mathbb{Z}_N$ denotes a hash function, $N$ the RSA modulus, $e$ the RSA public exponent and $d$ the RSA private exponent.}
\label{fig:dupless_protocol}
\end{figure}

DupLESS~\cite{dupless} allows clients to derive encryption keys for secure deduplication in cloud-based storage. Key derivation is performed in DupLESS by means of an interactive protocol between a client and a gateway based on RSA blind-signatures. The protocol is sketched in Figure~\ref{fig:dupless_protocol}. The client secret input is a file $M$, while the server secret input is the private exponent of an RSA key-pair. The corresponding public exponent is available to both parties. The client computes the hash of the file $M$ and blinds it with a random value $r$ that he raises to the public exponent $e$. He transmits the blinded hash value to the gateway. The gateway now signs the blinded value with its private exponent $d$. The gateway finally transmits the signed blinded hash back to the client. As $ed \equiv 1 \mod{\varphi(N)}$, we have that $y \equiv \left(hr^e \right)^d \equiv h^d r^{ed} \equiv h^d r \mod{N}$. The client can compute the $r^{-1} \mod{N}$, remove the blinding from $y$ and obtain the signed hash $h^d \mod{N}$. The client needs now to check the validity of the signature using the public exponent of the gateway $e$. If the signature is valid, the generated symmetric key will be the hash of the signed hash of the file $K = H(z) = H(h^d)$.

The benefits of such a key generation protocol are two-fold:
\begin{itemize}
    \item Since the protocol is oblivious, it ensures that the gateway does not learn any information about the file. On the other hand, this protocol enables the client to check the correctness of the computation performed by the gateway (i.e., verify the gateway's signature).
\item By involving the gateway in the key generation process, brute-force attacks on predictable messages (i.e., files) can be slowed down by rate-limiting key-generation requests to the gateway. %Notice that this scheme does not prevent a curious gateway from performing brute-force searches on predictable messages, acquiring the hash, and the corresponding key. %In this sense, the security offered by our scheme reduces to that of existing MLE schemes (cf. Section~\ref{subsec:security}).
\end{itemize}

\balance
\end{document}